%% file: main.tex
\documentclass[sigconf]{acmart}
\AtBeginDocument{%
  }

\setcopyright{acmlicensed}
\copyrightyear{2026}
\acmYear{2026}
\acmDOI{XXXXXXX.XXXXXXX}
\acmYear{2026}\copyrightyear{2026}
\setcopyright{cc}
\setcctype[4.0]{by}
\acmConference[EuroMLSys '26]{The 6th Workshop on Machine Learning and Systems}{April 27--30, 2026}{Edinburgh, Scotland Uk}
\acmBooktitle{The 6th Workshop on Machine Learning and Systems (EuroMLSys '26), April 27--30, 2026, Edinburgh, Scotland Uk}
\acmDOI{10.1145/3805621.3807610}
\acmISBN{979-8-4007-2605-7/26/04}

\newcommand{\xmark}{\ding{55}}   
\newcommand{\cmark}{\ding{51}}   
\usepackage{mdframed}
\usepackage{caption}
\usepackage{subcaption}

\newmdenv[
  linewidth=0.8pt,
  roundcorner=0pt,
  innerleftmargin=8pt,
  innerrightmargin=8pt,
  innertopmargin=6pt,
  innerbottommargin=6pt,
  skipabove=\medskipamount,
  skipbelow=\medskipamount
]{takeawaybox}

\newcounter{takeaway}
\newcommand{\takeaway}[2][]{%
  \refstepcounter{takeaway}%
  \begin{takeawaybox}
    \textbf{Takeaway~\#\thetakeaway:} \textit{#2}%
    \if\relax\detokenize{#1}\relax\else\label{#1}\fi
  \end{takeawaybox}
}

\input{preamble}

\begin{document}
\settopmatter{authorsperrow=4}
\title{LayoutBench: Performance Benchmarking of Cloud Storage Layouts for Multimedia Data}

\author{Debopam Sanyal}
\affiliation{
  \institution{Georgia Tech}
  \city{}
  \country{}}

\author{Hongjie Chen}
\affiliation{
  \institution{Dolby Labs}
  \city{}
  \country{}}

\author{Alexey Tumanov}
\affiliation{
  \institution{Georgia Tech}
  \city{}
  \country{}}

\author{Joshua Kimball}
\affiliation{
  \institution{Dolby Labs}
  \city{}
  \country{}
}

\renewcommand{\shortauthors}{Sanyal et al.}

\begin{abstract}
Modern multimedia machine learning workloads increasingly store large-scale datasets in cloud object storage services such as AWS S3.
How these samples are physically organized in storage (i.e., \emph{storage layout}) directly affects how quickly and cheaply they can be retrieved.
Yet the benchmarks used to guide storage decisions today focus on database engines and query processing, and none systematically evaluates how different storage layouts perform for multimedia data retrieval.
We present \textbf{LayoutBench}, the first benchmark designed to fill this gap. It evaluates three representative layout strategies: storing each sample as an individual object (L1), sequentially packing samples into tar archives (L2), and organizing samples as columns in Parquet files (L3).
We measure retrieval time, data transferred, and monetary cost using 11 queries of varying result-set sizes on ImageNet across six AWS EC2 instance configurations that span different network bandwidth and memory tiers.
Our experiments reveal that L2 achieves lower latency than L1 and L3 through connection reuse, but loses this advantage as retrieval sizes become very large.
L3 is the fastest for very large retrievals but transfers substantially more data across all query sizes due to row-group granularity, and requires significantly more memory.
Across all layouts, data transfer cost dominates total expenditure, with L3 costing an order of magnitude more than L1 or L2.
\end{abstract}

\ccsdesc[500]{Software and its engineering~Software design techniques}

\keywords{Storage Layouts, Benchmarks}

\maketitle

\section{Introduction}
Modern machine learning pipelines depend on large-scale datasets of unstructured data such as image, audio, and video~\cite{wanginternvid,nielsen2022mumin,yun2024web2code,sanyal2026klas,sanyal2023pareto}.
These datasets are increasingly stored in cloud object storage services such as AWS S3~\cite{zhang2023cdsben,antonopoulos2019socrates,arzhanov2025applying}, which serve as the primary persistence layer for training and inference workloads~\cite{li2024gaussdb,chen2022cloudjump,mao2024bytemq,wang2025diagnosing}.
Before any model can train, the relevant data samples must be retrieved from cloud storage and delivered to compute nodes, a process whose efficiency directly affects GPU utilization, pipeline throughput, and infrastructure cost.
How samples are grouped, packed, and organized in cloud storage what we term the \emph{storage layout}, fundamentally determines the retrieval performance and cost of this data loading step.
Yet, practitioners today must choose among diverse storage layout strategies with no benchmark to systematically compare their retrieval performance, data transfer overhead, and cost under realistic workloads.

Existing data systems benchmarks do not fill this gap.
Benchmarks for OnLine Transaction Processing (OLTP)~\cite{weng2024lauca,zhang2023cdsben}, OnLine Analytical Processing (OLAP)~\cite{lutsch2024benchmarking}, time-series databases~\cite{hao2021ts,khelifati2023tsm,wang2022timeunion}, and other workloads~\cite{kim2022m2bench,zhang2024hybench,pan2024survey,zhang2025cloudybench} measure query latency, throughput, and scalability under a fixed storage abstraction, assuming that the underlying data organization is either given or outside the scope of evaluation.
These benchmarks target structured, tabular data and do not capture the distinct access patterns of multimedia ML workloads, where individual data samples are large binary objects (e.g., JPEG images, audio files) retrieved using metadata predicates rather than relational joins.
As a result, the performance tradeoffs in storing and retrieving such data are governed by factors absent from existing benchmarks.

Prior work has studied related but distinct aspects of cloud storage.
Benchmarks such as COSBench~\cite{zheng2013cosbench} and CNSBench~\cite{merenstein2021cnsbench} evaluate storage services (e.g., throughput and latency of object stores) but do not vary the data layout within a given service.
LavaStore~\cite{wang2024lavastore} optimizes key-value storage layouts, but targets structured workloads rather than multimedia retrieval.
Zeng~et~al.~\cite{zeng2023empirical} compare columnar \emph{formats} (primarily Parquet vs.\ ORC) but focus on relational analytics rather than multimedia retrieval.
Pixels~\cite{bian2022pixels} optimizes storage layouts for relational data, and Delta Lake~\cite{armbrust2020delta,hai2023data} stores data as Parquet files with a co-located transaction log containing data-skipping statistics, but its primary concern is change tracking and ACID transactions rather than retrieval performance.
BigLake~\cite{levandoski2024biglake} unifies storage and analytics at scale but does not benchmark alternative layout strategies against one another.
On the ML systems side, WebDataset~\cite{aizman2019high} and MosaicML Streaming~\cite{mosaicml2022streaming} adopt sequential packing into shards, while Petastorm~\cite{petastormdocs} and Lance~\cite{pace2025lance} use columnar storage, yet none provides a controlled performance comparison of storage layout strategies for multimedia retrieval on cloud object storage with cost analysis.

We propose \textbf{LayoutBench}, a benchmark framework for evaluating cloud storage layouts for multimedia ML datasets.
LayoutBench targets \emph{data sample retrieval}: fetching specific subsets of a dataset from cloud storage based on predicate conditions.
Examples include retrieving training images of a particular class, loading audio segments that meet a size constraint, or selecting video clips matching resolution criteria.
Such retrieval operations are pervasive in production systems, from serving image results on platforms like Pinterest to loading audio tracks on Spotify~\cite{shiau2020shop,regan2023semi}.
Even with caching, indexing, and query optimization at higher layers, the organization of data at the storage layer directly affects retrieval latency, network transfer volume, and monetary cost~\cite{ji2024lbsc,mukherjee2023towards}.

This paper makes the following contributions:
\textbf{(1)}~We propose an extensible benchmark into which new storage layouts, cloud backends, and datasets can be plugged.
We validate it by implementing three representative layout strategies using AWS S3: individual object storage~(L1), sequential packing into tar archives~(L2), and columnar storage via Parquet~(L3), spanning the design space from per-sample access to structured columnar retrieval.
\textbf{(2)}~Through experiments on ImageNet~\cite{deng2009imagenet} with 11 representative retrieval queries across three dataset scales and six EC2 instance configurations, we provide an empirical characterization of the performance, data transfer, and cost tradeoffs of each layout.
Our findings reveal that each layout is constrained by a different resource (L1 by per-request latency, L2 by network bandwidth, and L3 by memory), that data transfer accounts for over 98\% of total cost, and that L2 offers the best latency-cost balance for image workloads.
These results provide actionable guidance for ML practitioners configuring their storage layers.

\section{Storage Layouts}
Every storage layout must manage two kinds of information: \emph{metadata} (attributes such as class label, file size, width, and height that are used to select samples, i.e., images) and the \emph{raw sample data} itself (i.e., image bytes).
The three layouts we benchmark differ in how they co-locate or separate these two concerns, and consequently in how retrieval is performed.

\begin{figure}[!ht]
  \centering
  \includegraphics[width=\columnwidth]{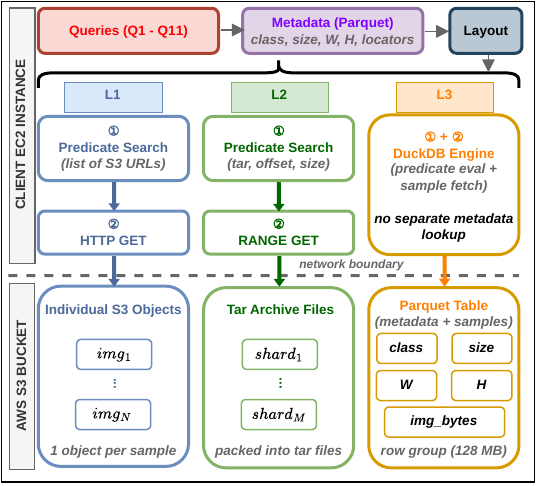}
  \captionsetup{skip=2pt}
  \caption{\textit{LayoutBench} retrieval architecture. The dashed
  line marks the network boundary between the client EC2
  instance and AWS S3. L1 and L2 perform a local metadata lookup
  (\ding{192}) followed by per-sample S3 requests (\ding{193}).
  L3 combines both steps in a single DuckDB query that internally
  issues S3 API calls. The bottom contrasts
  how each layout organizes data in S3: many individual objects
  (L1), fewer tar archives (L2), or a single co-located Parquet
  table (L3).}
  \label{fig:architecture}
\end{figure}

Figure~\ref{fig:architecture} illustrates the overall architecture.
In all layouts, data resides in AWS S3 buckets and a local client EC2 instance runs a query engine that operates in two phases: \ding{192}~a \emph{predicate search} that consults metadata to identify which samples satisfy a query's conditions, and \ding{193}~a \emph{retrieval step} that issues requests to S3 to fetch the matching samples.
L1 and L2 perform these two phases separately: the predicate search returns sample locators (S3 URLs for L1, byte-range triplets for L2), and the retrieval step issues one HTTP request per locator.
L3 combines both phases in a single DuckDB query that internally issues S3 API calls, requiring no separate metadata lookup.
The bottom of the figure contrasts how each layout organizes data in S3: many individual objects (L1), fewer tar archives (L2), or a co-located Parquet table (L3).
We describe each layout below.

\subsection{Layout 1: Individual Sample Storage}
Each sample is stored as an individual S3 object with a unique S3 URL.
The client first queries the local metadata table to obtain the URLs of matching samples, then issues one HTTP~GET request per URL to fetch each sample. 
Sec.\ref{appendix-sec-L1-sql} provides the query details for L1.

\medskip\noindent\textbf{\textit{Discussion:}}
This is the most primitive storage layout.
Existing literature~\cite{bian2022pixels,arzhanov2025applying} has noted that retrieving individual files incurs high Time-to-First-Byte (TTFB) latency per request.

\subsection{Layout 2: Sequential Sample (Tar) Storage}
Samples are sequentially packed into tar files.
The client-side metadata table additionally records each sample's tar file URL, byte offset, and size.
The client queries this table to obtain the (tar URL, offset, size) triplet for each matching sample, then issues one HTTP~RANGE~GET request per triplet, fetching only the relevant byte range from the corresponding tar file.
Sec.\ref{appendix-sec-L2-sql} provides the query details for L2.

\medskip\noindent\textbf{\textit{Discussion:}}
Other sequential formats such as WebDataset~\cite{aizman2019high,totten2021scaling,xu2022efficient} and TFRecord prioritize streaming efficiency over random access retrieval.
We select \textit{tar} because it supports byte-range access, making it suitable for predicate-driven retrieval.

\subsection{Layout 3: Samples in Columnar Storage}
Unlike L1 and L2, Layout~3 stores the content bytes of each sample as a BLOB in a Parquet column alongside the metadata columns.
The client issues SQL queries through a DuckDB connector, which internally performs predicate evaluation and sample fetching via low-level S3 API calls, i.e., no separate metadata lookup is needed.
To DuckDB, all samples appear to reside in a single large Parquet table.
Sec.\ref{appendix-sec-L3-sql} provides the complete SQL statements used in DuckDB.


\medskip\noindent\textbf{\textit{Discussion:}}
Other columnar formats (e.g., ORC, Lance) and query engines (e.g., Apache Spark) can also serve this role~\cite{zeng2023empirical}.
We select Parquet and DuckDB due to their widespread adoption.

\section{Experimental Configuration}
This section describes the configuration for datasets, queries, client instances and metrics.
Additional details are provided in Sec.\ref{appendix-sec-exp-details}.

\begin{table}[!ht]
    \centering
    \footnotesize
    \captionsetup{skip=2pt}
    \caption{
    Dataset statistics at the storage server and client.
    }
\resizebox{\linewidth}{!}{
    \begin{tabular}{@{}c@{\phantom{1}}c@{\phantom{1}}c@{\phantom{1}}c@{\phantom{1}}c@{\phantom{1}}c@{}}
        \toprule
        
        \multirow{2}{*}{\textbf{Dataset}} &
        \multirow{2}{*}{\textbf{Location}} &
        \multirow{2}{*}{\textbf{Attribute}} &
        \multicolumn{3}{c}{\textbf{Storage Layout Plan}} \\
        
        \cmidrule(r){4-6}
        & & & \textbf{L1} & \textbf{L2} & \textbf{L3} \\
        \midrule

        \multirow{4}{*}{mini}
        & \multirow{2}{*}{Server}
        & Number of Files & 12,756 & 23 & 22 \\
        & 
        & File Size & 1.39 GB & 1.38 GB & 1.36 GB \\

        \cmidrule(r){2-6}

        & Client
        & Index File Size & 553 KB & 810 KB & 0 \\

        \midrule

        \multirow{4}{*}{medium}
        & \multirow{2}{*}{Server}
        & Number of Files & 128,066 & 28 & 28 \\
        &
        & File Size & 13.91 GB & 13.76 GB & 13.57 GB \\

        \cmidrule(r){2-6}

        & Client
        & Index File Size & 4.4 MB & 6.8 MB & 0 \\

        \midrule

        \multirow{4}{*}{full}
        & \multirow{2}{*}{Server}
        & Number of Files & 1,281,167 & 276 & 274 \\
        &
        & File Size & 139.25 GB & 137.73 GB & 135.85 GB \\

        \cmidrule(r){2-6}

        & Client
        & Index File Size & 35.6 MB & 54.7 MB & 0 \\

        \bottomrule
    \end{tabular}
}
    \label{tab:data-statistics-server-client}
\end{table}

\begin{table*}[!ht]
\centering
\captionsetup{skip=2pt}
\caption{
Summary of the queries.  
Each query performs a predicate search on one or more of four conditions: Image Class (C), Size (S), Width (W), and Height (H).  
Queries Q1–Q6 use a single condition, while Q7–Q11 involve two or more conditions.
}
\resizebox{\linewidth}{!}{
\begin{tabular}{l l cccc l rrr}
\toprule
\multirow{2}{*}{\textbf{Q.}} &
\multirow{2}{*}{\textbf{Type}} &
\multicolumn{4}{c}{\textbf{Predicate}} &
\multicolumn{1}{c}{\multirow{2}{*}{\textbf{Query Description}}} &
\multicolumn{3}{c}{\textbf{Retrieval} -- \textbf{Num. of Files}} \\
\cmidrule(lr){3-6} \cmidrule(lr){8-10}
 & & \textbf{C} & \textbf{S} & \textbf{W} & \textbf{H} & & \textbf{mini} & \textbf{medium} & \textbf{full} \\
\midrule
Q1  & One Sample         & \checkmark & & & & one image of \textit{streetcar} & 1     & 1      & 1      \\
Q2  & Batch Fetch        & \checkmark & & & & N images of \textit{lemon} ($N=10 (mini), 100 (medium), 1000 (full)$) & 10    & 100    & 1,000  \\
Q3  & One Class          & \checkmark & & & & images of \textit{cello} & 13    & 130    & 1,300  \\
Q4  & Regex Match        & \checkmark & & & & images whose labels are prefixed with snake, i.e., LIKE \textit{\%snake} & 130   & 1,300  & 13,000 \\
Q5  & Fan-out            & \checkmark & & & & images of \textit{Old\_English\_sheepdog}, \textit{giant\_panda}, and \textit{water\_buffalo} & 39    & 390    & 3,900  \\
Q6  & File size Filter   & & \checkmark & & & images that are at least 500 KB & 86    & 801    & 7,837  \\
\midrule
Q7  & Resolution Filter  & & & \checkmark & \checkmark & images whose width and height are both at least 1024 pixels & 133   & 1,309  & 12,635 \\
Q8 & Cross-column       & & & \checkmark & \checkmark & images whose width / height ratio is greater than 1.5 & 2,311 & 23,006 & 229,710 \\
Q9  & Multi-label Filter & \checkmark & \checkmark & & & images of \textit{canoe} and \textit{starfish} that are smaller than 100 KB & 6     & 47     & 575    \\
Q10  & Selective Filter   & \checkmark & \checkmark & & & images of \textit{tiger\_cat} that are smaller than 100 KB & 1     & 27     & 294    \\
Q11 & All-column Filter & \checkmark & \checkmark & \checkmark & \checkmark & images of \textit{bagel} under 200 KB with width and height at least 512 pixels & 1 & 6 & 32 \\
\bottomrule
\end{tabular}
}
\label{tab:data-statistics-per-query}
\end{table*}

\begin{table}[!ht]
\centering
\normalsize
\captionsetup{skip=2pt}
\caption{Comparison of client EC2 instance types}
\resizebox{\linewidth}{!}{
\begin{tabular}{H l r r r r@{} H H H H H}
\toprule
Name       & API Name   & Network & Memory & vCPUs   & Hourly Cost & Arch. & Compute Family    & Linux Reserved & Linux Minimum & CoreMark Score \\
\midrule                                                                               
T3 Medium  & t3.medium  & Up to 5 Gigabit     & 4 GiB  & 2 vCPUs & \$0.0416    & x86   & General purpose   & \$0.0261       & \$0.0183      & 29{,}163.021 \\
C5 Large   & c5.large   & Up to 10 Gigabit    & 4 GiB  & 2 vCPUs & \$0.0850    & x86   & Compute optimized & \$0.0540       & \$0.0299      & 33{,}103.448 \\
C8GB Large & c8gb.large & Up to 20 Gigabit    & 4 GiB  & 2 vCPUs & \$0.1185    & arm   & Compute optimized &                & \$0.0311      & 60{,}163.947 \\
\midrule                                                                               
T3 Large   & t3.large   & Up to 5 Gigabit     & 8 GiB  & 2 vCPUs & \$0.0832    & x86   & General purpose   & \$0.0522       & \$0.0374      & 27{,}292.576 \\
T3 XLarge  & t3.xlarge  & Up to 5 Gigabit     & 16 GiB & 4 vCPUs & \$0.1664    & x86   & General purpose   & \$0.1043       & \$0.0744      & 53{,}932.584 \\
T3 2XLarge & t3.2xlarge & Up to 5 Gigabit     & 32 GiB & 8 vCPUs & \$0.3328    & x86   & General purpose   & \$0.2086       & \$0.0984      & 113{,}426.911 \\
\bottomrule
\end{tabular}
}
\label{tab:ec2-comparison}
\end{table}

\subsection{Datasets}
Our benchmarking is based on \textit{ImageNet-1K}, one of the most widely used image datasets in the computer vision community, whose training subset contains 1.28~million images across 1{,}000~classes~\cite{deng2009imagenet}.
From this subset we derive three datasets at increasing scale:
(1)~\textit{mini}, a 1\% random sample of each class;
(2)~\textit{medium}, a 10\% random sample; and
(3)~\textit{full}, the complete training subset.
Table~\ref{tab:data-statistics-server-client} summarizes the data statistics for each (dataset, layout) combination.

Under L1, the number of files and total size correspond directly to the number of samples and raw dataset size, since each image is stored as an individual S3 object.
Under L2 and L3, samples are packed into tar or Parquet files, with a per-file size limit of 64~MB for mini and 512~MB for medium and full.
Note that the metadata information is stored in the client under L1 and L2.

\subsection{Queries}
We design 11 queries that reflect practical retrieval patterns in ML pipelines, where training typically involves filtering and selecting specific samples rather than scanning entire datasets~\cite{hambardzumyan2022deep}.
Each query is a predicate search over one or more of four metadata attributes: image Class~(C), file Size~(S), Width~(W), and Height~(H).
The queries fall into two groups, as summarized in Table~\ref{tab:data-statistics-per-query}.
Atomic queries (Q1-Q6) apply a single predicate.
Q1-Q5 filter on class using increasingly broad conditions: a single sample (Q1), a fixed batch (Q2), an entire class (Q3), a regex match across multiple classes (Q4), and a fan-out over three named classes (Q5).
Q6, on the other hand, filters solely based on file size.
The result sets for these queries range from one file (Q1) to 13{,}000~files (Q4) on the full dataset, spanning four orders of magnitude.

Composite queries (Q7-Q11) combine two or more predicates.
Q7 filters on both width and height; Q8 computes a cross-column ratio (width / height~$>$~1.5), requiring evaluation over all width and height values; Q9 and Q10 combine class membership with a size constraint at different selectivities; and Q11 involves all four attributes.
Result sets again span a wide range, from 32~files (Q11) to 229{,}710~files (Q8) on the full dataset.
This design ensures coverage across three axes: predicate type (membership, comparison, regex, cross-column computation), predicate count (one through four), and result set size (1 to 229{,}710~files).

\subsection{Client Instances}
We primarily evaluate on three types of EC2 instances: \texttt{t3.medium}, \texttt{c5.large}, and \texttt{c8gb.large}, due to their differences in network bandwidth.
When memory becomes a bottleneck, we opt for three other EC2 instances, \texttt{t3.large}, \texttt{t3.xlarge}, and \texttt{t3.2xlarge}, which provide additional memory.
Their specifications and prices are provided in Table~\ref{tab:ec2-comparison}.

\subsection{Metrics}
We measure three metrics:
(1)~\textit{End-to-end Retrieval Time} ($T$): the elapsed time from the start of query execution on the client to the completion of all sample retrieval from S3.
(2)~\textit{Data Transferred} ($D$): the total bytes transferred from S3 to the client.
(3)~\textit{Cost Estimate} ($C$): the sum of S3 data transfer cost and EC2 rental cost, computed from AWS pricing~\cite{aws_ec2_on_demand_pricing} (detailed in \S\ref{subsec:cost}).

\section{Results}
We present experimental results analyzing the three storage layouts across multiple dimensions: retrieval latency, data transfer efficiency, scalability, caching behavior, and cost.
To account for run-to-run variability, experiments were repeated five times for the mini dataset and three times for the medium dataset, with error bars showing standard deviations.
The full dataset was run once due to its scale.

\begin{figure}[!ht]
\centering
\begin{subfigure}[b]{\columnwidth}
  \centering
  \includegraphics[width=\columnwidth]{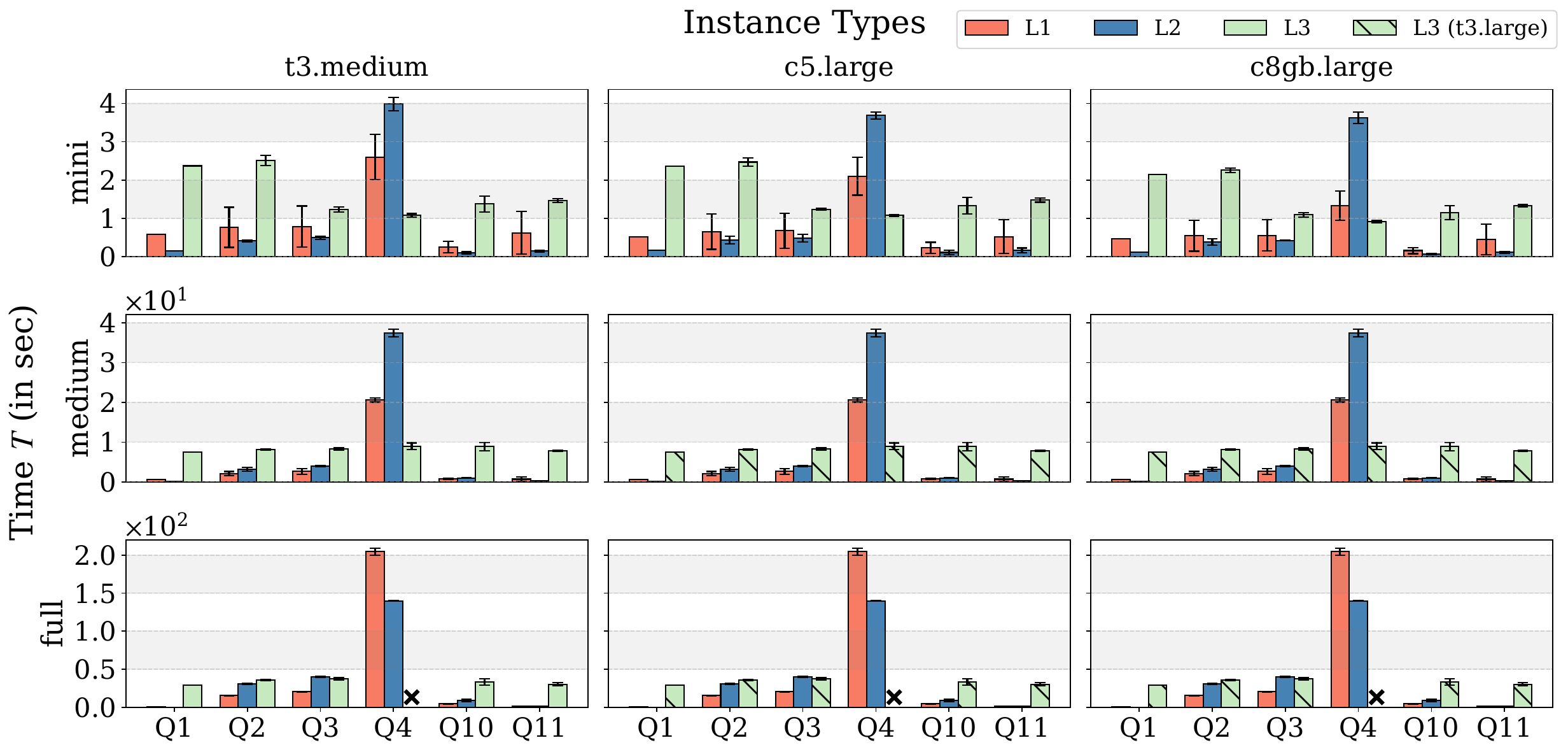}
  \caption{Queries: Q1-Q4 \& Q10-Q11.}
  \label{fig:total-time-datasetsize-by-instancetypes-part1}
\end{subfigure}
\begin{subfigure}[b]{\columnwidth}
  \centering
  \includegraphics[width=\columnwidth]{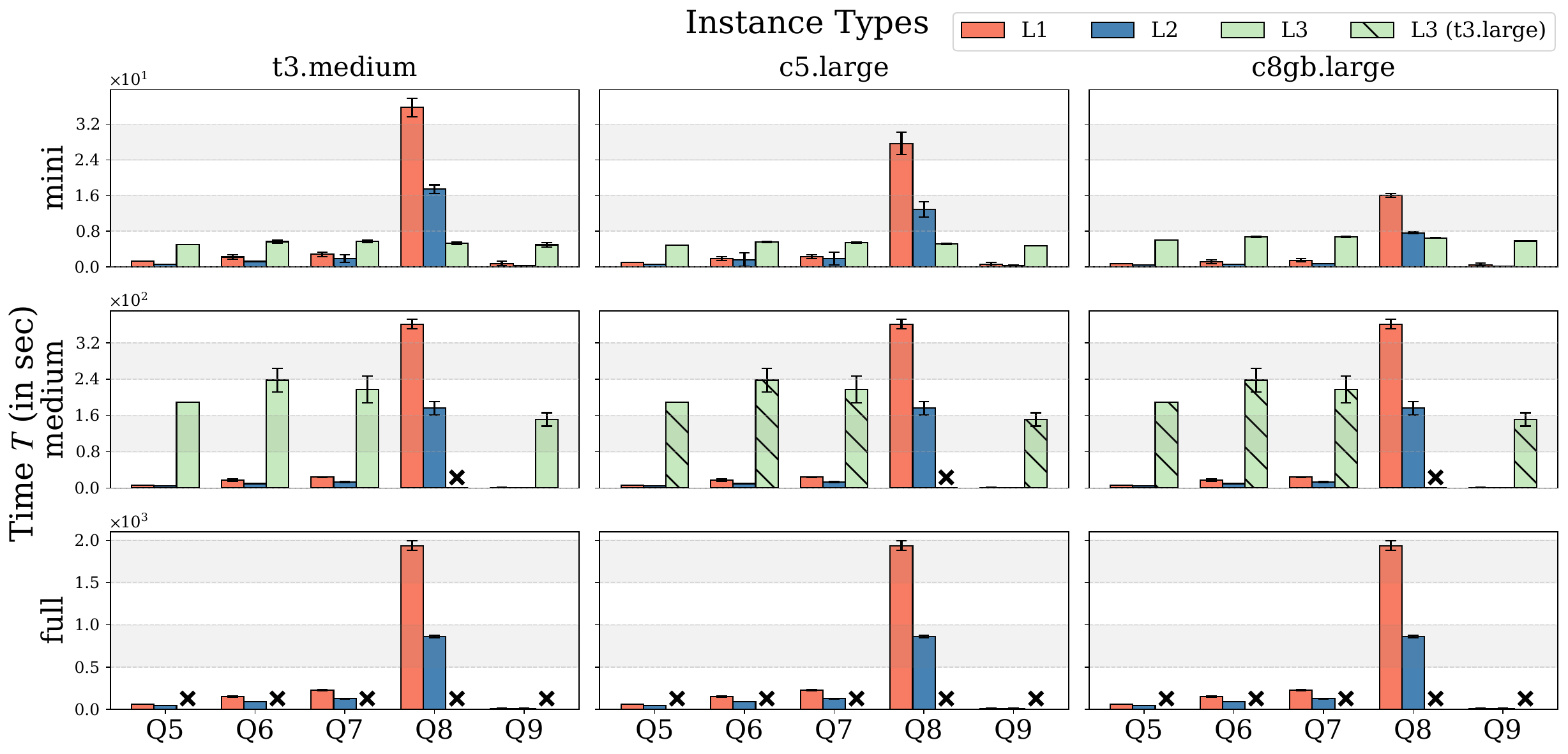}
  \caption{Queries: Q5-Q9.}
  \label{fig:total-time-datasetsize-by-instancetypes-part2}
\end{subfigure}
\captionsetup{skip=2pt}
\caption{
End-to-end retrieval time in seconds of three layouts (L1, L2, L3) on the EC2 instance types in Table~\ref{tab:ec2-comparison} (per column) for three dataset scales (per row: mini, medium, full).
}
\label{fig:total-time-datasetsize-by-instancetypes}
\end{figure}


\subsection{Layout Performance Comparison}
\label{subsec:performance}
Fig.~\ref{fig:total-time-datasetsize-by-instancetypes} presents the end-to-end retrieval time across all three layouts (L1, L2, L3) for the 11~queries on four EC2 instance types (\texttt{t3.medium}, \texttt{c5.large}, \texttt{c8gb.large}, \texttt{t3.large}) and three dataset scales (mini, medium, full).
Fig.~\ref{fig:total-time-datasetsize-by-instancetypes-part1} covers Q1-Q4 and Q10-Q11, while Fig.~\ref{fig:total-time-datasetsize-by-instancetypes-part2} covers Q5-Q9.
Each row corresponds to a dataset scale and each column to an instance type; the y-axis scales differ across rows.
A hatched bar (L3 with \texttt{t3.large}) appears on the medium and full rows where the default instances lacked sufficient memory for L3.
We highlight three key takeaways.

For queries with very small result sets (1-13 files), L2 is faster than L1.
In Figure~\ref{fig:total-time-datasetsize-by-instancetypes-part1}, mini row, L2 bars are consistently shorter than L1 bars for Q1, Q2, Q3, Q10, and Q11 across all instance types.
This is because L2's RANGE GET requests to tar files incur lower per-request overhead than L1's individual object GET requests at this \textit{extremely small} scale.

As the retrieval becomes slightly bigger, however, L1 slightly outperforms L2 for these same queries (Q1-Q3, Q10, Q11), visible in the medium and full rows of Figure~\ref{fig:total-time-datasetsize-by-instancetypes-part1}.
When samples are distributed across a few hundreds of tar files, L2 cannot reuse connections effectively, and its advantage disappears.
This is most visible on Q4: on mini (130 files), L1 is clearly faster than L2 across all instance types, but on full (13{,}000 files) L2 outperforms L1.
The same pattern holds in Figure~\ref{fig:total-time-datasetsize-by-instancetypes-part2}: L2 achieves visibly lower latency than L1 for Q5-Q8, particularly on the medium and full datasets, where the four queries return hundreds to thousands of files.
This advantage stems from L2's use of HTTP RANGE~GET requests to a small number of tar files, which enables TCP connection reuse and reduces per-request overhead compared to L1's individual GET requests.
\takeaway{\textbf{L2 outperforms L1 for large retrievals, but L1 is faster for smaller retrievals.}}

L3 incurs a visible baseline overhead relative to L1 and L2, even for the smallest queries.
In Fig.~\ref{fig:total-time-datasetsize-by-instancetypes-part1}, mini row, Q1 retrieves a single sample yet L3 takes 2-3 seconds while L1 and L2 complete in under 1 second.
Q10 and Q11 show the same pattern, with L3 at 1-2 seconds versus sub-second for L1 and L2.
For mid-range result sets (Q5-Q7, Q9 in Fig.~\ref{fig:total-time-datasetsize-by-instancetypes-part2}), L3 incurs substantially higher latency than L1 and L2 across all dataset scales and instance types.
However, Q8 and Q4 are two notable exceptions: despite being the most data-intensive queries, L3 is faster than both L1 and L2.
For example, on mini, L3 completes Q8 in approximately 5~seconds while L2 takes over 15 seconds and L1 takes over 30~seconds.
This reversal occurs because Q8's large result set (2{,}311~images on mini) causes L1 and L2's per-request overheads to dominate, while L3's row group reads become relatively efficient.
On the full dataset, L3 could not complete Q5-Q9 on the smaller instances, as indicated by $\times$ marks in the figures; larger-memory instances are required.
\takeaway{\textbf{L3 carries a baseline overhead that makes it slow for small and mid-range retrievals, but it is the fastest for the largest retrievals.}}

It is clear from Fig.~\ref{fig:total-time-datasetsize-by-instancetypes} that the dominant factor in retrieval time is the number of samples returned, not the number or type of predicates involved.
On the medium and full datasets, Q8 exhibits the highest latency for L1 and L2 because it returns the largest retrievals on medium and full (see Table~\ref{tab:data-statistics-per-query}).
In contrast, Q11 involves all four predicate columns (class, size, width, height) yet produces small bars across all layouts because it returns only 32~files on the full dataset.
Similarly, Q9 (575~files) and Q10 (294~files) complete quickly despite combining two predicates each.
This ordering is consistent across all three instance types in both sub-figures.
\takeaway{\textbf{Result set (or retrieval) size, not predicate complexity, determines retrieval time.}}

\subsection{Data Transfer Scalability}
\label{subsec:data-transfer}
\begin{figure}[!ht]
\centering
\begin{subfigure}[t]{0.49\linewidth}
  \centering
  \includegraphics[width=\linewidth]{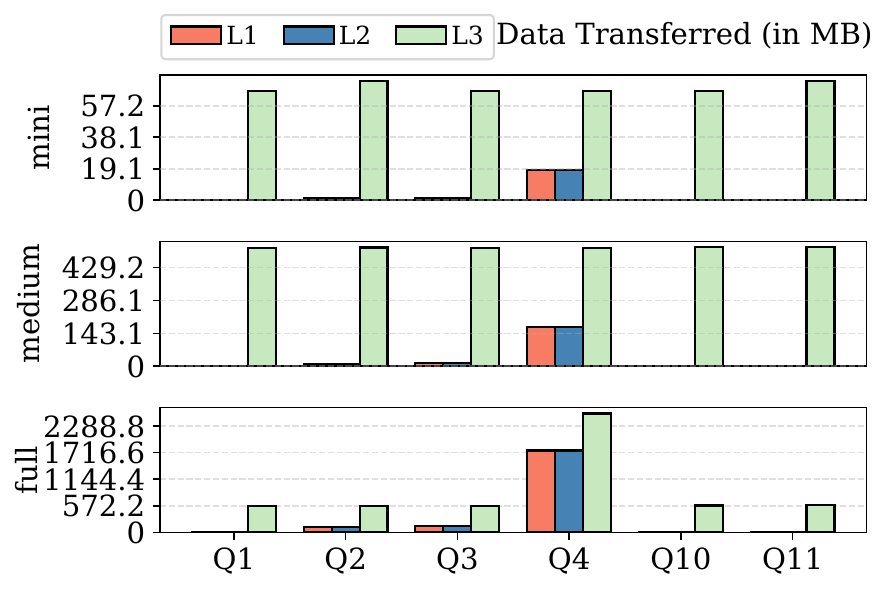}
  \caption{
  \footnotesize
  Queries: Q1-Q4 \& Q10-Q11.
  }
  \label{fig:bytestransferred-per-dataset-part1}
\end{subfigure}\hfill
\begin{subfigure}[t]{0.49\linewidth}
  \centering
  \includegraphics[width=\linewidth]{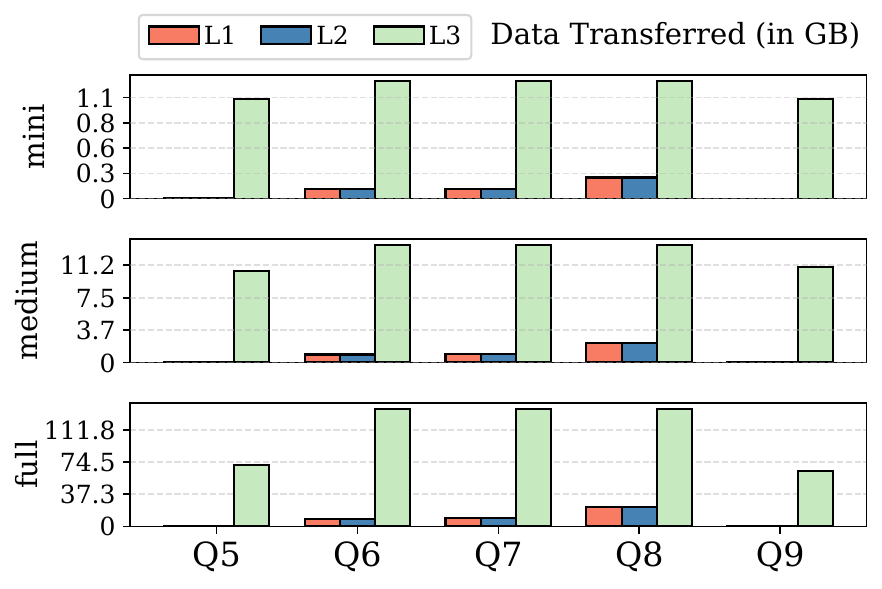}
  \caption{
  \footnotesize
  Queries: Q5-Q9.
  }
  \label{fig:bytestransferred-per-dataset-part2}
\end{subfigure}
\captionsetup{skip=2pt}
\caption{Data transferred in (a) megabytes and (b) gigabytes of three layouts (L1, L2, L3) for three dataset scales (per row: mini, medium, full) across all 11 queries.}
\label{fig:bytestransferred-per-dataset}
\end{figure}

Fig.~\ref{fig:bytestransferred-per-dataset} shows the total data transferred from S3 to the client for each query across the three layouts and dataset scales.
Fig.~\ref{fig:bytestransferred-per-dataset-part1} covers Q1-Q4 and Q10-Q11 with units in~MB, while Fig.~\ref{fig:bytestransferred-per-dataset-part2} covers Q5-Q9 with units in~GB.
Each row corresponds to a dataset scale (mini, medium, full); the y-axis scales differ across rows.
In Fig.~\ref{fig:bytestransferred-per-dataset}, we report results using the smallest instance that could complete the given query in our experiments.

L1 and L2 transfer nearly identical bytes for most queries across both sub-figures, as both layouts fetch only the requested image bytes plus minimal protocol overhead.
For Q1-Q3 and Q10-Q11 in Fig.~\ref{fig:bytestransferred-per-dataset-part1}, the L1 and L2 bars are barely visible next to L3.
Q4 is an exception: because it returns up to 13{,}000 files on the full dataset, L1 and L2 transfer visible amounts of data ($>$1{,}700~MB), though still less than L3.
Data transfer under L1 / L2 scales linearly with result set size.
For example, Q8 transfers approximately 0.3~GB, 3~GB, and 30~GB under L1 / L2 for mini, medium, and full datasets respectively (Fig.~\ref{fig:bytestransferred-per-dataset-part2}), consistent with the roughly 10$\times$ scaling in sample counts across the corresponding dataset sizes (see Table~\ref{tab:data-statistics-per-query}).
\takeaway[takeaway:l1-l2-data-transfer]{\textbf{L1 and L2 achieve near-optimal data transfer, fetching close to the minimum necessary bytes.}}

L3 transfers more data than L1 and L2 across all queries in both sub-figures, because DuckDB's Parquet reader fetches entire row groups containing the requested samples, reading adjacent rows even when only a subset matches the predicate.
The overhead ratio depends on the result set size.
For highly selective queries, the gap is extreme: Q1 (a single sample) transfers $\sim$ 57~MB under L3 on the mini dataset versus $<$ 1~MB under L1 / L2 (Fig.~\ref{fig:bytestransferred-per-dataset-part1}).
Similar patterns hold for Q2, Q3, Q10, and Q11, where L3 bars dominate the chart while L1 / L2 bars are negligible.
For queries returning large result sets, the gap reduces.
For example, on Q4 and Q8, which return the largest result sets, the gap between L3 and L1 / L2 narrows relative to other queries but remains substantial (L3 transfers roughly 3-4$\times$ more data than L1 / L2 across all dataset scales on Q8).
For all other large queries (Q5-Q7, Q9), L3 bars reach the top of the y-axis while L1 / L2 bars remain near the baseline, indicating that L3 transfers an order of magnitude more data than L1 / L2.
L3 incurs higher data transfer volume than L1 / L2 primarily due to row group granularity and the fixed overhead of Parquet metadata reading.
\takeaway[takeaway:l3-data-transfer]{\textbf{L3 transfers substantially more data due to row group granularity and fixed metadata reading cost, across queries and dataset scales.}}

\subsection{Layout Scalability to Instance Types}
\label{subsec:scalability}
\begin{figure}[!ht]
\centering
\begin{subfigure}[b]{\columnwidth}
  \centering
  \includegraphics[width=\columnwidth]{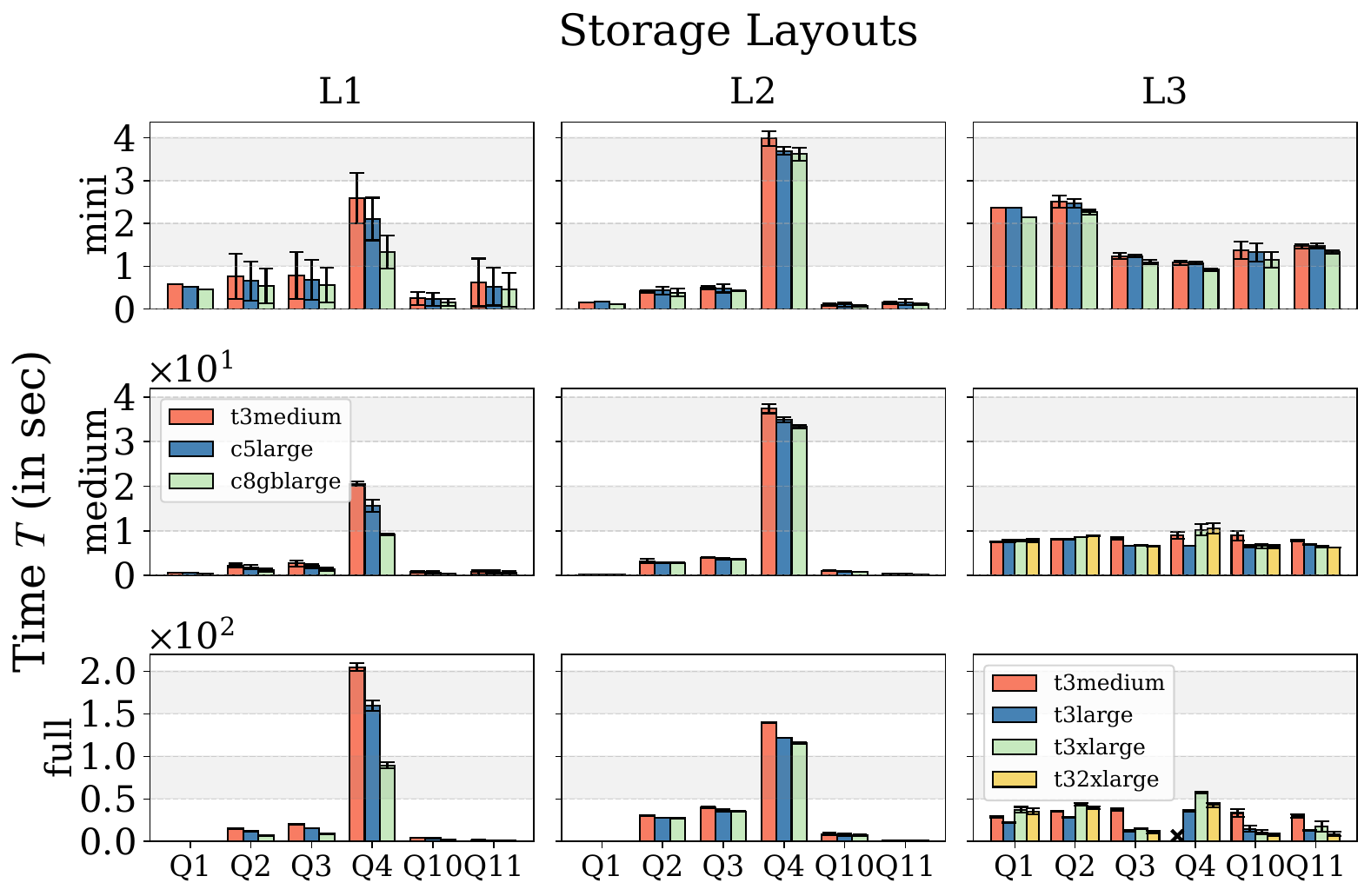}
  \caption{Queries: Q1-Q4 \& Q10-Q11.}
  \label{fig:total-time-layoutplans-by-datasetsize-part1}
\end{subfigure}
\begin{subfigure}[b]{\columnwidth}
  \centering
  \includegraphics[width=\columnwidth]{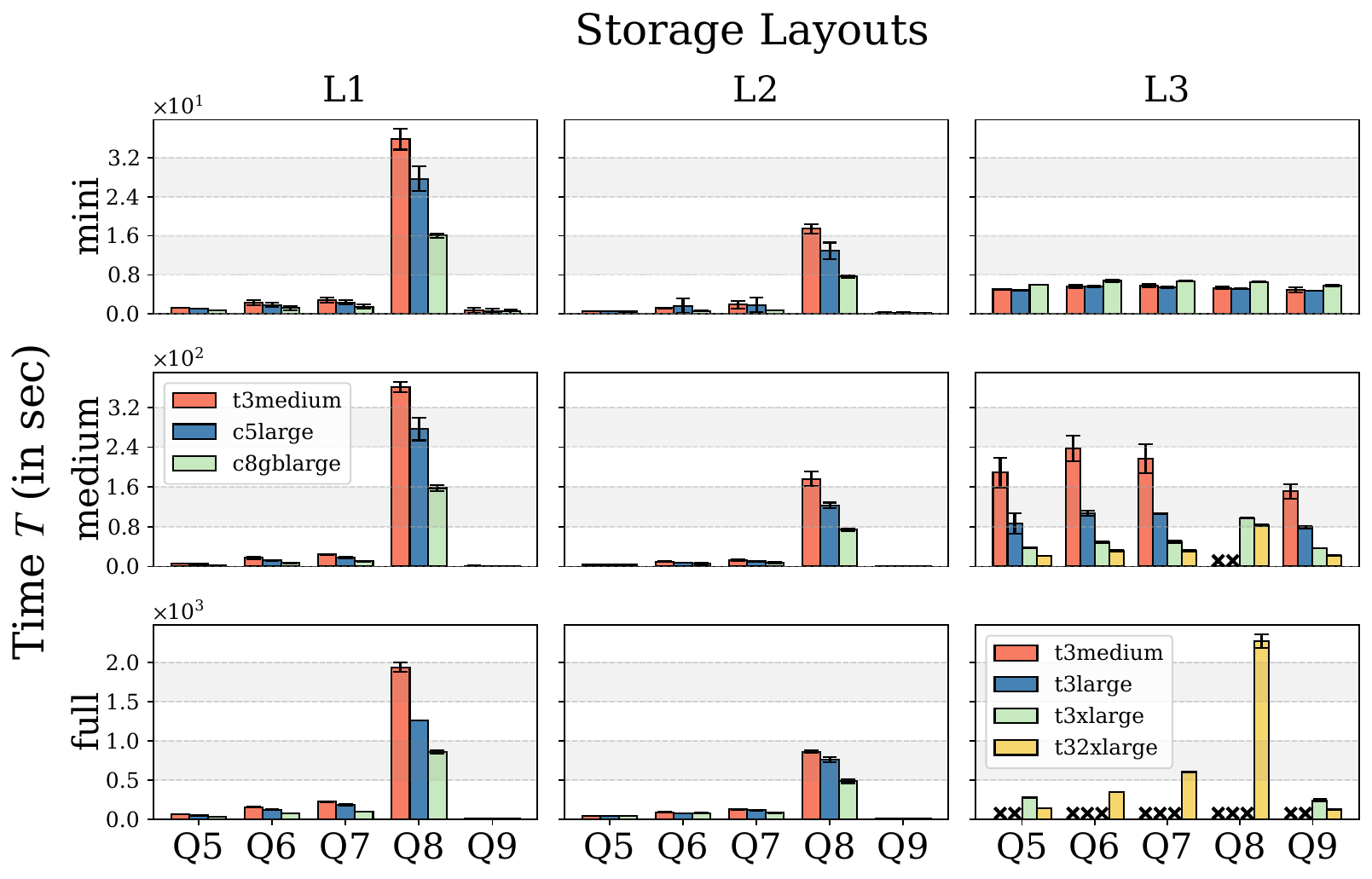}
  \caption{Queries: Q5-Q9.}
  \label{fig:total-time-layoutplans-by-datasetsize-part2}
\end{subfigure}
\captionsetup{skip=2pt}
\caption{
End-to-end retrieval time $T$ in seconds of different EC2 instance types on layouts (per column: L1, L2, L3) for three dataset scales (per row: mini, medium, full).
}
\label{fig:total-time-layoutplans-by-datasetsize}
\end{figure}

Fig.~\ref{fig:total-time-layoutplans-by-datasetsize} examines how each layout scales across different EC2 instance configurations.
Unlike Fig.~\ref{fig:total-time-datasetsize-by-instancetypes}, which groups bars by layout within each instance type, this figure groups bars by instance type within each layout (one column per layout), making it easier to see how a given layout responds to increased resources.
Each row corresponds to a dataset scale (mini, medium, full).
Fig.~\ref{fig:total-time-layoutplans-by-datasetsize-part1} covers Q1-Q4 and Q10-Q11; Fig.~\ref{fig:total-time-layoutplans-by-datasetsize-part2} covers Q5-Q9.
For L1 and L2, bars are colored by the three bandwidth-tier instances (\texttt{t3.medium}, \texttt{c5.large}, \texttt{c8gb.large}).
For L3 on full, the legend changes to memory-tier instances (\texttt{t3.medium}, \texttt{t3.large}, \texttt{t3.xlarge}, \texttt{t3.2xlarge}), as the bandwidth-tier instances with 4~GB memory could not complete execution at this scale ($\times$ marks in the figures).

In the L1 column, higher-bandwidth instances visibly reduce retrieval time for data-intensive queries.
The effect is clearest for Q4 and Q8 across all dataset scales.
In Fig.~\ref{fig:total-time-layoutplans-by-datasetsize-part1}, L1 column, full row, Q4 retrieval time drops from $\sim$200~seconds on \texttt{t3.medium} to $\sim$85~seconds on \texttt{c8gb.large}, a reduction of more than 50\%.
A similar pattern holds for Q8 in Fig.~\ref{fig:total-time-layoutplans-by-datasetsize-part2}: on the full dataset, \texttt{t3.medium} takes $\sim$2{,}000~seconds while \texttt{c8gb.large} completes in $\sim$800~seconds.
The L2 column shows a similar trend, though the reductions are less dramatic because L2 already benefits from connection reuse to fewer tar files.
For queries with small result sets (e.g., Q1, Q10, Q11 in Fig.~\ref{fig:total-time-layoutplans-by-datasetsize-part1}), bars are uniformly short across all instance types in both L1 and L2, confirming that bandwidth is not the bottleneck when little data is transferred.
\takeaway{\textbf{L1 and L2 are both bandwidth-bound: a faster network reduces latency for large retrievals, but has little effect on small retrievals.}}

The L3 column shows a different pattern.
On the mini dataset, L3 runs on the same three bandwidth-tier instances as L1 and L2, and the bars remain similar across instance types in both figures, indicating that network bandwidth is not L3's bottleneck.
On the full dataset, the smaller instances could not complete execution, requiring the largest instance instead, for all queries Q5-Q9 in Fig.~\ref{fig:total-time-layoutplans-by-datasetsize-part2}.
Among the memory-tier instances on the full dataset, the bars for smaller queries (e.g., Q1-Q3 in Fig.~\ref{fig:total-time-layoutplans-by-datasetsize-part1}, L3 column, full row) are similar across instance types, suggesting diminishing returns once DuckDB has sufficient memory to operate.
For larger queries, on the full dataset (Fig.~\ref{fig:total-time-layoutplans-by-datasetsize-part2}, L3 column, full row), only the largest memory-tier instance completed execution, with most smaller instances marked as failed.
\takeaway{\textbf{L3 is memory-bound rather than bandwidth-bound, and requires progressively larger instances for larger retrievals.}}

These scalability patterns have direct implications for instance selection.
For large retrievals under L1 or L2, investing in \texttt{c8gb.large} (\$0.1185/hr; Table~\ref{tab:ec2-comparison}) yields meaningful latency reductions over \texttt{t3.medium} (\$0.0416/hr).
For selective queries, \texttt{t3.medium} provides comparable performance, as the bars in Fig.~\ref{fig:total-time-layoutplans-by-datasetsize-part1} confirm.
For L3 on the full dataset, the required \texttt{t3.2xlarge} (32~GiB, \$0.3328/hr) costs 8$\times$ more per hour than the \texttt{t3.medium} that suffices for L1 and L2.

\subsection{Cold vs.\ Warm Runs}
To understand caching effects, we compare cold runs (first execution after instance launch) against warm runs (subsequent executions).
Table~\ref{tab:cold-warm-run-ratio} reports the ratio of cold-run to warm-run retrieval time, averaged across all queries, on \texttt{t3.medium}.
A ratio above 1.0 indicates a cold-start penalty (cold runs are slower), while a ratio below 1.0 indicates that cold runs are faster than warm runs.

\begin{table}[!ht]
\centering
\begin{minipage}[b]{0.50\columnwidth}
\centering
\normalsize
\small
\captionsetup{skip=2pt}
\caption{Cold-to-warm runtime ratio on \texttt{t3.medium} across layouts and dataset scales.}
\label{tab:cold-warm-run-ratio}
\resizebox{1\linewidth}{!}{
\begin{tabular}{lccc}
\toprule
Dataset & \textbf{L1} & \textbf{L2} & \textbf{L3} \\
\midrule
mini   & 1.186 & 0.987 & 1.056 \\
medium & 0.978 & 0.980 & 1.078 \\
full   & 0.962 & 0.987 & 1.073 \\
\bottomrule
\end{tabular}
}
\end{minipage}
\begin{minipage}[b]{0.49\columnwidth}
\centering
\normalsize
\small
\captionsetup{skip=2pt}
\caption{Cost breakdown on \textit{t3.medium} (excluding Q8) across layouts.}
\label{tab:cost-breakdown-t3-medium}
\resizebox{1\linewidth}{!}{
\begin{tabular}{@{}lccc@{}}
\toprule
 & \textbf{L1} & \textbf{L2} & \textbf{L3} \\
\midrule
$C_{\text{network}}$ & \$0.795 & \$0.795 & \$9.183 \\
$C_{\text{compute}}$   & \$0.010 & \$0.005 & \$0.019 \\
\midrule
$C (total)$        & \$0.805 & $\mathbf{\$0.800}$ & \$9.202 \\
\bottomrule
\end{tabular}
}
\end{minipage}
\end{table}

L2 exhibits the most stable behavior, with ratios between 0.980 and 0.987 across all dataset scales, i.e., essentially no difference between cold and warm runs.
This is consistent with L2's retrieval mechanism: each RANGE~GET request is independent, and there is little client-side state to cache between runs.
L1 shows a more varied pattern.
On the mini dataset, L1's ratio is 1.186, indicating a notable 19\% cold-start penalty, likely reflecting TCP connection establishment and DNS resolution costs that are amortized in warm runs through connection pooling.
However, on the medium (0.978) and full (0.962) datasets, L1's ratios fall below 1.0, meaning warm runs are slightly \emph{slower} than cold runs.
This counterintuitive result may arise from increased resource contention on the client during warm runs or from the negligible role of connection-level caching when the number of GET requests grows to hundreds of thousands.

L3 consistently shows ratios above 1.0 (1.056-1.078), indicating a modest 5-8\% cold-start penalty across all dataset scales.
This overhead is attributable to DuckDB's one-time initialization costs: Parquet footer reading, metadata parsing, and query plan compilation.
On warm runs, DuckDB benefits from cached metadata and pre-compiled execution plans, as well as cached columnar statistics (min/max values per row group) that enable more efficient predicate pushdown.
Unlike L1, L3's cold-start penalty is consistent across dataset scales, suggesting that the overhead is dominated by fixed initialization costs rather than per-sample factors.
\takeaway{\textbf{Caching effects are modest and dependent on layout and dataset scale.}}

\subsection{Cost Estimates}
\label{subsec:cost}
We model the total cost $C$ for executing a set of queries as the sum of two components: $C = C_{\text{network}} + C_{\text{compute}}$, where $C_{\text{network}} = \$0.09 \times D$ is the S3 data transfer cost for $D$~GB of data downloaded, and $C_{\text{compute}} = r \times T$ is the EC2 rental cost at hourly rate $r$ for total retrieval time $T$ (in hours).
The hourly rates for each instance type are listed in Table~\ref{tab:ec2-comparison}~\cite{aws_ec2_on_demand_pricing}.
We present the cost breakdown on the medium dataset using \texttt{t3.medium}, summed across all queries except Q8, in Table~\ref{tab:cost-breakdown-t3-medium} .
We exclude Q8 because L3 runs out of memory on \texttt{t3.medium}, as shown in Fig.~\ref{fig:total-time-datasetsize-by-instancetypes-part2}.

Across all three layouts, $C_{\text{network}}$ accounts for over 98\% of total cost.
Compute cost $C_{\text{compute}}$ is negligible by comparison.
This confirms that for cloud-based multimedia retrieval, the bill is determined almost entirely by how many bytes leave S3, not by how long the client runs.
L1 and L2 incur identical transfer costs (\$0.795) because both fetch only the requested image bytes (Takeaway~\ref{takeaway:l1-l2-data-transfer}).
L2 achieves the lowest total cost (\$0.800) thanks to its slightly faster retrieval, which reduces $C_{\text{compute}}$ to half that of L1 (\$0.005 vs.\ \$0.010).
L3's total cost (\$9.202) is 11.5$\times$ higher than L2, driven almost entirely by its excessive data transfer due to row group granularity (Takeaway~\ref{takeaway:l3-data-transfer}).
This means the storage layout choice is not just a performance decision but a significant economic one: running queries under L3 costs an order of magnitude more than under L1 or L2.
\takeaway{\textbf{Data transfer cost dominates total expenditure, and L3's row group overhead carries a steep price.}}

\begin{table}[!ht]
\centering
\footnotesize
\normalsize
\captionsetup{skip=2pt}
\caption{Performance characterization by layout (L1, L2, L3). Each layout exhibits distinct resource constraints that dominate as the retrieval sizes become progressively larger.
}
\label{tab:bottleneck-summary}
\begin{tabular}{lccc}
\toprule
\textbf{Characteristic} & \textbf{L1} & \textbf{L2} & \textbf{L3} \\
\midrule
Primary bottleneck & Latency & Bandwidth & Memory \\
Connection overhead & High & Low & N/A \\
Predicate pushdown & \xmark & \xmark & \cmark \\
Min. instance (full) & t3.medium & t3.medium & t3.2xlarge \\
Scales with bandwidth & \cmark & \cmark & \xmark \\
\bottomrule
\end{tabular}
\end{table}

\textbf{\textit{Overall Performance Characteristics}}:
Table~\ref{tab:bottleneck-summary} summarizes the overall performance characterization of each layout.
L1 is primarily bottlenecked by per-request latency, L2 by network bandwidth, and L3 by memory.
L2 achieves a good latency-cost balance for ImageNet-1K: it matches or outperforms L1 across most medium to moderately large retrieval sizes, runs on the cheapest instance type, and incurs the lowest total cost.
L3 is viable only for very large query retrievals where fast retrieval via predicate pushdown justifies its higher memory and data transfer costs.

\section{Limitations and Conclusion}
In our study, we evaluate only the image modality; performance tradeoffs may differ for larger binary objects such as video or audio, where per-sample sizes are orders of magnitude greater.
Additionally, we benchmark read-only retrieval and do not measure write or ingest paths, which may favor different layouts.
Finally, our experiments use a single cloud provider and a single query engine for L3; results may vary with other storage backends or engines.

We presented \textbf{LayoutBench}, the first benchmark for systematically evaluating how cloud storage layouts affect multimedia data retrieval performance and cost.
Our experiments across three layouts, 11 queries, three dataset scales, and six EC2 instance types reveal that L2 offers a good latency-cost balance for images, achieving low latency through connection reuse while incurring low total cost.
LayoutBench is extensible by design: one can easily plug in new layouts, storage backends, datasets, and modalities.
Future work includes benchmarking write and ingest paths, evaluating hybrid layouts that combine L2's I/O efficiency with L3's predicate pushdown, and extending to text and vector retrieval workloads.

\bibliographystyle{ACM-Reference-Format}
\bibliography{main}

\clearpage
\newpage
\appendix

\section{SQL Statements for Retrieval in L1}
\label{appendix-sec-L1-sql}
The SQL statements \ref{enumerate-L1-q1}-\ref{enumerate-L1-q11} are used in L1 experiments.

\begin{enumerate}[label=\textbf{(Q\arabic*)}]

\item \label{enumerate-L1-q1} One Sample Preview: Get one image of \textit{streetcar}.
\begin{lstlisting}
SELECT filename
FROM read_parquet('{CLI_METADATA_L1_PATH}')
WHERE human_label = 'streetcar'
LIMIT 1;
\end{lstlisting}

\item Batch Fetch: Get N images of \textit{lemon}.
\begin{lstlisting}
SELECT filename
FROM read_parquet('{CLI_METADATA_L1_PATH}')
WHERE human_label = 'lemon'
LIMIT N;
\end{lstlisting}

\item One Class: Get all images of \textit{cello}.
\begin{lstlisting}
SELECT filename
FROM read_parquet('{CLI_METADATA_L1_PATH}')
WHERE human_label = 'cello';
\end{lstlisting}

\item Regex Match: Get images whose labels are LIKE \textit{\%snake}.
\begin{lstlisting}
SELECT filename
FROM read_parquet('{CLI_METADATA_L1_PATH}')
WHERE human_label LIKE '%snake';
\end{lstlisting}

\item Fan-out: Get images of \textit{Old\_English\_sheepdog}, \textit{giant\_panda}, and \textit{water\_buffalo}.
\begin{lstlisting}
SELECT filename
FROM read_parquet('{CLI_METADATA_L1_PATH}')
WHERE human_label IN ('Old_English_sheepdog','giant_panda','water_buffalo');
\end{lstlisting}

\item File Size Filter: Get images that are at least 500 KB.
\begin{lstlisting}
SELECT filename
FROM read_parquet('{CLI_METADATA_L1_PATH}')
WHERE filesize_bytes >= 500 * 1024;
\end{lstlisting}

\item Resolution Filter: Get images whose width and height are both at least 1024 pixels.
\begin{lstlisting}
SELECT filename
FROM read_parquet('{CLI_METADATA_L1_PATH}')
WHERE width >= 1024
  AND height >= 1024;
\end{lstlisting}

\item Cross-column Predicate: Get images whose width / height ratio is greater than 1.5.
\begin{lstlisting}
SELECT filename,
       width,
       height
FROM read_parquet('{CLI_METADATA_L1_PATH}')
WHERE CAST(width AS DOUBLE) / CAST(height AS DOUBLE) > 1.5;
\end{lstlisting}

\item Multi-label Filter: Get images of \textit{canoe} and \textit{starfish} that are smaller than 100 KB.
\begin{lstlisting}
SELECT filename
FROM read_parquet('{CLI_METADATA_L1_PATH}')
WHERE human_label IN ('canoe', 'starfish')
  AND filesize_bytes < 100 * 1024;
\end{lstlisting}

\item Selective Filter: Get images of \textit{tiger\_cat} that are smaller than 100 KB.
\begin{lstlisting}
SELECT filename
FROM read_parquet('{CLI_METADATA_L1_PATH}')
WHERE human_label = 'tiger_cat'
  AND filesize_bytes < 100 * 1024;
\end{lstlisting}

\item \label{enumerate-L1-q11} Composite Filter: Get images of \textit{bagel} whose file size is smaller than 200 KB and width and height are both at least 512 pixels.
\begin{lstlisting}
SELECT filename
FROM read_parquet('{CLI_METADATA_L1_PATH}')
WHERE human_label = 'bagel'
  AND width >= 512
  AND height >= 512
  AND filesize_bytes < 200 * 1024;
\end{lstlisting}

\end{enumerate}

The returned filenames are used to infer their S3 URLs in order to fetch the image samples.
\begin{lstlisting}
L1: Initialize S3 transfer with multithreading

For each file URL in parallel:
    Download file to local directory
\end{lstlisting}

\section{SQL Statements for Retrieval in L2}
\label{appendix-sec-L2-sql}
The SQL statements \ref{enumerate-L2-q1}-\ref{enumerate-L2-q11} are used in L2.

\begin{enumerate}[label=\textbf{(Q\arabic*)}]

\item \label{enumerate-L2-q1} One Sample Preview: Get one image of \textit{streetcar}.
\begin{lstlisting}
SELECT filename, shard, data_offset, data_size
FROM read_parquet('{CLI_METADATA_L2_PATH}')
WHERE human_label = 'streetcar'
LIMIT 1;
\end{lstlisting}

\item Batch Fetch: Get N images of \textit{lemon}.
\begin{lstlisting}
SELECT filename, shard, data_offset, data_size
FROM read_parquet('{CLI_METADATA_L2_PATH}')
WHERE human_label = 'lemon'
LIMIT N;
\end{lstlisting}

\item One Class: Get all images of \textit{cello}.
\begin{lstlisting}
SELECT filename, shard, data_offset, data_size
FROM read_parquet('{CLI_METADATA_L2_PATH}')
WHERE human_label = 'cello';
\end{lstlisting}

\item Regex Match: Get images whose labels are LIKE \textit{\%snake}.
\begin{lstlisting}
SELECT filename, shard, data_offset, data_size
FROM read_parquet('{CLI_METADATA_L2_PATH}')
WHERE human_label LIKE '%snake';
\end{lstlisting}

\item Fan-out: Get images of \textit{Old\_English\_sheepdog}, \textit{giant\_panda}, and \textit{water\_buffalo}.
\begin{lstlisting}
SELECT filename, shard, data_offset, data_size
FROM read_parquet('{CLI_METADATA_L2_PATH}')
WHERE human_label IN ('Old_English_sheepdog','giant_panda','water_buffalo');
\end{lstlisting}

\item File Size Filter: Get images that are at least 500 KB.
\begin{lstlisting}
SELECT filename, shard, data_offset, data_size
FROM read_parquet('{CLI_METADATA_L2_PATH}')
WHERE filesize_bytes >= 500 * 1024;
\end{lstlisting}

\item Resolution Filter: Get images whose width and height are both at least 1024 pixels.
\begin{lstlisting}
SELECT filename, shard, data_offset, data_size
FROM read_parquet('{CLI_METADATA_L2_PATH}')
WHERE width >= 1024
  AND height >= 1024;
\end{lstlisting}

\item Cross-column Predicate: Get images whose width / height ratio is greater than 1.5.
\begin{lstlisting}
SELECT filename, shard, data_offset, data_size, width, height
FROM read_parquet('{CLI_METADATA_L2_PATH}')
WHERE CAST(width AS DOUBLE) / CAST(height AS DOUBLE) > 1.5;
\end{lstlisting}

\item Multi-label Filter: Get images of \textit{canoe} and \textit{starfish} that are smaller than 100 KB.
\begin{lstlisting}
SELECT filename, shard, data_offset, data_size
FROM read_parquet('{CLI_METADATA_L2_PATH}')
WHERE human_label IN ('canoe', 'starfish')
  AND filesize_bytes < 100 * 1024;
\end{lstlisting}

\item Selective Filter: Get images of \textit{tiger\_cat} that are smaller than 100 KB.
\begin{lstlisting}
SELECT filename, shard, data_offset, data_size
FROM read_parquet('{CLI_METADATA_L2_PATH}')
WHERE human_label = 'tiger_cat'
  AND filesize_bytes < 100 * 1024;
\end{lstlisting}

\item \label{enumerate-L2-q11} Composite Filter: Get images of \textit{bagel} whose file size is smaller than 200 KB and width and height are both at least 512 pixels.
\begin{lstlisting}
SELECT filename, shard, data_offset, data_size
FROM read_parquet('{CLI_METADATA_L2_PATH}')
WHERE human_label = 'bagel'
  AND width >= 512
  AND height >= 512
  AND filesize_bytes < 200 * 1024;
\end{lstlisting}

\end{enumerate}

The returned filenames, along with shard, data offset, and data size, are used to fetch the corresponding byte ranges from S3.

\begin{lstlisting}
L2: Initialize S3 transfer with multithreading
    For each file in parallel:
        Download the byte range corresponding to shard, data_offset and data_size
\end{lstlisting}

\section{SQL Statements for Retrieval in L3}
\label{appendix-sec-L3-sql}
The SQL statements \ref{enumerate-L3-q1}-\ref{enumerate-L3-q11} are used in DuckDB.
\begin{enumerate}[label=\textbf{(Q\arabic*)}]
\item \label{enumerate-L3-q1} One Sample Preview: Get one image of \textit{streetcar}.
\begin{lstlisting}
SELECT filename, image_bytes
FROM read_parquet('{Parquet_S3_PATH}')
WHERE label = 'streetcar'
LIMIT 1
\end{lstlisting}

\item Batch Fetch: Get N images of \textit{lemon}.
\begin{lstlisting}
SELECT filename, image_bytes
FROM read_parquet('{Parquet_S3_PATH}')
WHERE label = 'lemon'
LIMIT N
\end{lstlisting}

\item One Class: Get all images of \textit{cello}.
\begin{lstlisting}
SELECT filename, image_bytes
FROM read_parquet('{Parquet_S3_PATH}')
WHERE label = 'cello'
\end{lstlisting}

\item Regex Match: Get images whose labels are LIKE \textit{\%snake}.
\begin{lstlisting}
SELECT filename, image_bytes
FROM read_parquet('{Parquet_S3_PATH}')
WHERE label LIKE '%snake';
\end{lstlisting}

\item Fan-out: Get images of \textit{Old\_English\_sheepdog}, \textit{giant\_panda}, and \textit{water\_buffalo}.
\begin{lstlisting}
SELECT filename, image_bytes
FROM read_parquet('{Parquet_S3_PATH}')
WHERE label
IN ('Old_English_sheepdog','giant_panda','water_buffalo')
\end{lstlisting}

\item File Size Filter: Get images that are at least 500 KB.
\begin{lstlisting}
SELECT filename, image_bytes
FROM read_parquet('{Parquet_S3_PATH}')
WHERE filesize_bytes >= 500 * 1024;
\end{lstlisting}

\item Resolution Filter: Get images whose width and height are both at least 1024 pixels.
\begin{lstlisting}
SELECT filename, image_bytes
FROM read_parquet('{Parquet_S3_PATH}')
WHERE width >= 1024
  AND height >= 1024
\end{lstlisting}

\item Cross-column Predicate: Get images whose width / height ratio is greater than 1.5. 
\begin{lstlisting}
SELECT filename, image_bytes
FROM read_parquet('{Parquet_S3_PATH}')
WHERE CAST(width AS DOUBLE) / CAST(height AS DOUBLE) > 1.5
\end{lstlisting}

\item Multi-label Filter: Get images of \textit{canoe} and \textit{starfish} that are smaller than 100 KB.
\begin{lstlisting}
SELECT filename, image_bytes
FROM read_parquet('{Parquet_S3_PATH}')
WHERE label IN ('canoe', 'starfish')
  AND filesize_bytes < 100 * 1024
\end{lstlisting}

\item Selective Filter: Get images of \textit{tiger\_cat} that are smaller than 100 KB.
\begin{lstlisting}
SELECT filename, image_bytes
FROM read_parquet('{Parquet_S3_PATH}')
WHERE label = 'tiger_cat'
  AND filesize_bytes < 100 * 1024
\end{lstlisting}

\item \label{enumerate-L3-q11} Composite Filter: Get images of \textit{bagel} whose file size is smaller than 200 KB and width and height are both at least 512 pixels.
\begin{lstlisting}
SELECT filename, image_bytes
FROM read_parquet('{Parquet_S3_PATH}')
WHERE label = 'bagel'
  AND width >= 512
  AND height >= 512
  AND filesize_bytes < 200 * 1024
\end{lstlisting}
\end{enumerate}
Notably, a separate download is not needed, as the image bytes are fetched directly within the SQL statements.

\section{Additional Experimental Details}
\label{appendix-sec-exp-details}

\noindent\textbf{\textit{Cold and Warm Runs.}}
After launching an instance, we iterate over all 11 queries and execute twice for each query, where the first execution corresponds to a cold run since there is no cache for the fetched samples.
The second execution corresponds to a warm run, where samples may be cached in the system.

\noindent\textbf{\textit{Number of Threads.}}
To fully utilize all resources provided by the client EC2 instances, we set the upper bound on the number of threads to 64 for the t3 family, 128 for \texttt{c5.large}, and 256 for \texttt{c8gb.large} under L1 and L2.
In all cases, the upper bound is sufficiently large that the number of spawned threads does not hit the limit.
Under L3, we configure the number of threads through DuckDB and set it to 32 by default.
When experimenting with the full dataset, we adjust the number of threads to avoid out-of-memory errors. 
Specifically, we set it as 8 for Q6 and Q9, 4 for Q7, and 2 for Q4.

\noindent\textbf{\textit{Storage on EC2 Clients.}}
The storage provisioned on EC2 clients must be large enough to accommodate downloaded data.
By default, we allocate 8\,GB for the mini dataset and 32\,GB for the medium and large datasets.
For larger queries, we increase the storage capacity up to 64\,GB as needed.


\end{document}

%% file: preamble.tex
\usepackage{listings}
\usepackage{enumitem}

\usepackage{amsmath}
\usepackage{pifont}
\usepackage{comment}
\usepackage{graphicx}
\usepackage{multirow}
\usepackage{booktabs}
\usepackage{caption}
\usepackage{capt-of}  
\usepackage{array}
\usepackage{hyperref}
\usepackage{xcolor}
\usepackage{lipsum}

\usepackage{color}
\usepackage{array}
\usepackage{colortbl}

\usepackage{tikz}

\newcolumntype{H}{>{\setbox0=\hbox\bgroup}c<{\egroup}@{}}

\definecolor{thedarkblue}{RGB}{0,0,120} 
\definecolor{mygreen}{RGB}{0,100,0} 
\definecolor{mydarkblue}{rgb}{0,0.08,0.45} 
\definecolor{darkblue}{rgb}{0,0.08,180}
\definecolor{orange}{HTML}{E6550D}
\colorlet{TufteRed}{red!80!black}

\definecolor{theblue}{RGB}{0,0,180}
\colorlet{thered}{TufteRed}

\providecommand{\eat}[1]{\ignorespaces}
\providecommand{\addressed}[1]{\ignorespaces}
\providecommand{\todoeat}[1]{\ignorespaces}
\providecommand{\supp}[1]{\ignorespaces} 

\DeclareMathOperator{\hugeE}{\mbox{\huge\raise-0.3ex\hbox{E}}}
\DeclareMathOperator{\p}{\mathbb{P}}
\DeclareMathOperator{\hugep}{\mbox{\huge\raise-0.3ex\hbox{$\p$}}}



%% file: main.bib
@inproceedings{deng2009imagenet,
  title={Imagenet: A large-scale hierarchical image database},
  author={Deng, Jia and Dong, Wei and Socher, Richard and Li, Li-Jia and Li, Kai and Fei-Fei, Li},
  booktitle={2009 IEEE conference on computer vision and pattern recognition},
  pages={248--255},
  year={2009},
  organization={Ieee}
}

@article{lutsch2024benchmarking,
  title={Benchmarking analytical query processing in intel SGXv2},
  author={Lutsch, Adrian and El-Hindi, Muhammad and Heinrich, Matthias and Ritter, Daniel and Istv{\u{A}}{\k{A}}n, Zsolt and Binnig, Carsten},
  journal={arXiv preprint arXiv:2403.11874},
  year={2024}
}

@article{pan2024survey,
  title={Survey of vector database management systems},
  author={Pan, James Jie and Wang, Jianguo and Li, Guoliang},
  journal={The VLDB Journal},
  volume={33},
  number={5},
  pages={1591--1615},
  year={2024},
  publisher={Springer}
}

@inproceedings{hao2021ts,
  title={Ts-benchmark: A benchmark for time series databases},
  author={Hao, Yuanzhe and Qin, Xiongpai and Chen, Yueguo and Li, Yaru and Sun, Xiaoguang and Tao, Yu and Zhang, Xiao and Du, Xiaoyong},
  booktitle={2021 IEEE 37th International Conference on Data Engineering (ICDE)},
  pages={588--599},
  year={2021},
  organization={IEEE}
}

@article{zhang2023cdsben,
  title={Cdsben: Benchmarking the performance of storage services in cloud-native database system at bytedance},
  author={Zhang, Jiashu and Jiang, Wen and Tang, Bo and Ma, Haoxiang and Cao, Lixun and Jiang, Zhongbin and Nie, Yuanyuan and Wang, Fan and Zhang, Lei and Liang, Yuming},
  journal={Proceedings of the VLDB Endowment},
  volume={16},
  number={12},
  pages={3584--3596},
  year={2023},
  publisher={VLDB Endowment}
}

@article{kim2022m2bench,
  title={M2bench: a database benchmark for multi-model analytic workloads},
  author={Kim, Bogyeong and Koo, Kyoseung and Enkhbat, Undraa and Kim, Sohyun and Kim, Juhun and Moon, Bongki},
  journal={Proceedings of the VLDB Endowment},
  volume={16},
  number={4},
  pages={747--759},
  year={2022},
  publisher={VLDB Endowment}
}

@article{zhang2024hybench,
  title={HyBench: A new benchmark for HTAP databases},
  author={Zhang, Chao and Li, Guoliang and Lv, Tao},
  journal={Proceedings of the VLDB Endowment},
  volume={17},
  number={5},
  pages={939--951},
  year={2024},
  publisher={VLDB Endowment}
}

@article{khelifati2023tsm,
  title={TSM-bench: benchmarking time series database systems for monitoring applications},
  author={Khelifati, Abdelouahab and Khayati, Mourad and Dign{\"o}s, Anton and Difallah, Djellel and Cudr{\'e}-Mauroux, Philippe},
  journal={Proceedings of the VLDB Endowment},
  volume={16},
  number={11},
  pages={3363--3376},
  year={2023},
  publisher={VLDB Endowment}
}

@article{weng2024lauca,
  title={Lauca: A workload duplicator for benchmarking transactional database performance},
  author={Weng, Siyang and Wang, Qingshuai and Qu, Luyi and Zhang, Rong and Cai, Peng and Qian, Weining and Zhou, Aoying},
  journal={IEEE Transactions on Knowledge and Data Engineering},
  volume={36},
  number={7},
  pages={3180--3194},
  year={2024},
  publisher={IEEE}
}

@inproceedings{antonopoulos2019socrates,
  title={Socrates: The new sql server in the cloud},
  author={Antonopoulos, Panagiotis and Budovski, Alex and Diaconu, Cristian and Hernandez Saenz, Alejandro and Hu, Jack and Kodavalla, Hanuma and Kossmann, Donald and Lingam, Sandeep and Minhas, Umar Farooq and Prakash, Naveen and others},
  booktitle={Proceedings of the 2019 International Conference on Management of Data},
  pages={1743--1756},
  year={2019}
}

@inproceedings{merenstein2021cnsbench,
  title={$\{$CNSBench$\}$: A cloud native storage benchmark},
  author={Merenstein, Alex and Tarasov, Vasily and Anwar, Ali and Bhagwat, Deepavali and Lee, Julie and Rupprecht, Lukas and Skourtis, Dimitris and Yang, Yang and Zadok, Erez},
  booktitle={19th USENIX Conference on File and Storage Technologies (FAST 21)},
  pages={263--276},
  year={2021}
}

@article{zeng2023empirical,
  title={An empirical evaluation of columnar storage formats},
  author={Zeng, Xinyu and Hui, Yulong and Shen, Jiahong and Pavlo, Andrew and McKinney, Wes and Zhang, Huanchen},
  journal={Proceedings of the VLDB Endowment},
  volume={17},
  number={2},
  pages={148--161},
  year={2023},
  publisher={VLDB Endowment}
}

@inproceedings{zheng2013cosbench,
  title={Cosbench: Cloud object storage benchmark},
  author={Zheng, Qing and Chen, Haopeng and Wang, Yaguang and Zhang, Jian and Duan, Jiangang},
  booktitle={Proceedings of the 4th ACM/SPEC International Conference on Performance Engineering},
  pages={199--210},
  year={2013}
}

@article{hambardzumyan2022deep,
  title={Deep lake: A lakehouse for deep learning},
  author={Hambardzumyan, Sasun and Tuli, Abhinav and Ghukasyan, Levon and Rahman, Fariz and Topchyan, Hrant and Isayan, David and McQuade, Mark and Harutyunyan, Mikayel and Hakobyan, Tatevik and Stranic, Ivo and others},
  journal={arXiv preprint arXiv:2209.10785},
  year={2022}
}

@inproceedings{aizman2019high,
  title={High performance I/O for large scale deep learning},
  author={Aizman, Alex and Maltby, Gavin and Breuel, Thomas},
  booktitle={2019 IEEE International Conference on Big Data (Big Data)},
  pages={5965--5967},
  year={2019},
  organization={IEEE}
}

@article{armbrust2020delta,
  title={Delta lake: high-performance ACID table storage over cloud object stores},
  author={Armbrust, Michael and Das, Tathagata and Sun, Liwen and Yavuz, Burak and Zhu, Shixiong and Murthy, Mukul and Torres, Joseph and Van Hovell, Herman and Ionescu, Adrian and {\L}uszczak, Alicja and others},
  journal={Proceedings of the VLDB Endowment},
  volume={13},
  number={12},
  pages={3411--3424},
  year={2020},
  publisher={VLDB Endowment}
}

@inproceedings{bian2022pixels,
  title={Pixels: An efficient column store for cloud data lakes},
  author={Bian, Haoqiong and Ailamaki, Anastasia},
  booktitle={2022 IEEE 38th International Conference on Data Engineering (ICDE)},
  pages={3078--3090},
  year={2022},
  organization={IEEE}
}

@inproceedings{levandoski2024biglake,
  title={BigLake: BigQuery's evolution toward a multi-cloud lakehouse},
  author={Levandoski, Justin and Casto, Garrett and Deng, Mingge and Desai, Rushabh and Edara, Pavan and Hottelier, Thibaud and Hormati, Amir and Johnson, Anoop and Johnson, Jeff and Kurzyniec, Dawid and others},
  booktitle={Companion of the 2024 International Conference on Management of Data},
  pages={334--346},
  year={2024}
}

@article{hai2023data,
  title={Data lakes: A survey of functions and systems},
  author={Hai, Rihan and Koutras, Christos and Quix, Christoph and Jarke, Matthias},
  journal={IEEE Transactions on Knowledge and Data Engineering},
  volume={35},
  number={12},
  pages={12571--12590},
  year={2023},
  publisher={IEEE}
}

@article{wang2024lavastore,
  title={LavaStore: ByteDance's Purpose-Built, High-Performance, Cost-Effective Local Storage Engine for Cloud Services},
  author={Wang, Hao and Ou, Jiaxin and Zhao, Ming and Qiu, Sheng and Jiao, Yizheng and Wang, Yi and Mao, Qizhong and Yang, Zhengyu and Liu, Yang and Zhang, Jianshun and others},
  journal={Proceedings of the VLDB Endowment},
  volume={17},
  number={12},
  pages={3799--3812},
  year={2024},
  publisher={VLDB Endowment}
}

@misc{aws_ec2_on_demand_pricing,
  title        = {Amazon EC2 On-Demand Pricing},
  author       = {{Amazon Web Services, Inc.}},
  howpublished = {\url{https://aws.amazon.com/ec2/pricing/on-demand/}},
  note         = {Accessed: 2026-02-22},
  year         = {2026},
}

@online{arzhanov2025applying,
  author       = {Alexander Arzhanov and Ilya Isaev and Roy Allela},
  title        = {Applying Data Loading Best Practices for ML Training with Amazon S3 Clients},
  year         = {2025},
  url          = {https://aws.amazon.com/blogs/machine-learning/applying-data-loading-best-practices-for-ml-training-with-amazon-s3-clients/},
  organization = {Amazon Web Services},
  note         = {Accessed: 2026-02-22 AWS Machine Learning Blog},
}

@online{totten2021scaling,
  author       = {Jordan Totten and Shane Hansen},
  title        = {Scaling deep learning workloads with PyTorch / XLA and Cloud TPU VM},
  year         = {2021},
  url          = {https://cloud.google.com/blog/topics/developers-practitioners/scaling-deep-learning-workloads-pytorch-xla-and-cloud-tpu-vm},
  organization = {Google Cloud},
  note         = {Accessed: 2026-02-22 Google Cloud Blog, Developers \& Practitioners},
}

@online{xu2022efficient,
  author       = {Xiang Xu and Rajesh Thallam},
  title        = {Efficient PyTorch training with Vertex AI},
  year         = {2022},
  url          = {https://cloud.google.com/blog/products/ai-machine-learning/efficient-pytorch-training-with-vertex-ai},
  organization = {Google Cloud},
  note         = {Accessed: Google Cloud Blog, AI \& Machine Learning},  
}

@inproceedings{wang2022timeunion,
  title={Timeunion: An efficient architecture with unified data model for timeseries management systems on hybrid cloud storage},
  author={Wang, Zhiqi and Shao, Zili},
  booktitle={Proceedings of the 2022 International Conference on Management of Data},
  pages={1418--1432},
  year={2022}
}

@article{li2024gaussdb,
  title={Gaussdb: A cloud-native multi-primary database with compute-memory-storage disaggregation},
  author={Li, Guoliang and Tian, Wengang and Zhang, Jinyu and Grosman, Ronen and Liu, Zongchao and Li, Sihao},
  journal={Proceedings of the VLDB Endowment},
  volume={17},
  number={12},
  pages={3786--3798},
  year={2024},
  publisher={VLDB Endowment}
}

@article{chen2022cloudjump,
  title={CloudJump: optimizing cloud databases for cloud storages},
  author={Chen, Zongzhi and Yang, Xinjun and Li, Feifei and Cheng, Xuntao and Hu, Qingda and Miao, Zheyu and Xie, Rongbiao and Wu, Xiaofei and Wang, Kang and Song, Zhao and others},
  journal={Proceedings of the VLDB Endowment},
  volume={15},
  number={12},
  pages={3432--3444},
  year={2022},
  publisher={VLDB Endowment}
}

@inproceedings{mao2024bytemq,
  title={Bytemq: A cloud-native streaming data layer in bytedance},
  author={Mao, Yancan and Yin, Ruohang and Lei, Liyuan and Ye, Peng and Zou, Shengfu and Tang, Shizheng and Guo, Yunzhe and Yuan, Ye and Yu, Xiaochen and Wan, Bo and others},
  booktitle={Proceedings of the 2024 ACM Symposium on Cloud Computing},
  pages={774--791},
  year={2024}
}

@inproceedings{zhang2025cloudybench,
  title={CloudyBench: A testbed for a comprehensive evaluation of cloud-native databases},
  author={Zhang, Chao and Li, Guoliang and Liu, Leyao and Lv, Tao and Fan, Ju},
  booktitle={2025 IEEE 41st International Conference on Data Engineering (ICDE)},
  pages={1--13},
  year={2025},
  organization={IEEE}
}

@inproceedings{wang2025diagnosing,
  title={Diagnosing and resolving cloud platform instability with multi-modal rag llms},
  author={Wang, Yifan and Birman, Kenneth P},
  booktitle={Proceedings of the 5th Workshop on Machine Learning and Systems},
  pages={139--147},
  year={2025}
}

@misc{mosaicml2022streaming,
    author = {The Mosaic ML Team},
    title = {streaming},
    year = {2022},
    howpublished = {\url{<https://github.com/mosaicml/streaming/>}},
}

@article{pace2025lance,
  title={Lance: Efficient random access in columnar storage through adaptive structural encodings},
  author={Pace, Weston and She, Chang and Xu, Lei and Jones, Will and Lockett, Albert and Wang, Jun and Shah, Raunak},
  journal={arXiv preprint arXiv:2504.15247},
  year={2025}
}

@misc{petastormdocs,
  title        = {Petastorm Documentation},
  howpublished = {\url{https://petastorm.readthedocs.io/en/latest/index.html}},
  note         = {Accessed: 2026-02-23},
  year         = {2022},
  organization = {Uber Technologies, Inc.},
}

@inproceedings{wanginternvid,
  title={InternVid: A Large-scale Video-Text Dataset for Multimodal Understanding and Generation},
  author={Wang, Yi and He, Yinan and Li, Yizhuo and Li, Kunchang and Yu, Jiashuo and Ma, Xin and Li, Xinhao and Chen, Guo and Chen, Xinyuan and Wang, Yaohui and others},
  booktitle={The Twelfth International Conference on Learning Representations}
}

@inproceedings{nielsen2022mumin,
  title={Mumin: A large-scale multilingual multimodal fact-checked misinformation social network dataset},
  author={Nielsen, Dan S and McConville, Ryan},
  booktitle={Proceedings of the 45th international ACM SIGIR conference on research and development in information retrieval},
  pages={3141--3153},
  year={2022}
}

@article{yun2024web2code,
  title={Web2code: A large-scale webpage-to-code dataset and evaluation framework for multimodal llms},
  author={Yun, Sukmin and Thushara, Rusiru and Bhat, Mohammad and Wang, Yongxin and Deng, Mingkai and Wang, Jinhong and Tao, Tianhua and Li, Junbo and Li, Haonan and Nakov, Preslav and others},
  journal={Advances in neural information processing systems},
  volume={37},
  pages={112134--112157},
  year={2024}
}

@inproceedings{shiau2020shop,
  title={Shop the look: Building a large scale visual shopping system at pinterest},
  author={Shiau, Raymond and Wu, Hao-Yu and Kim, Eric and Du, Yue Li and Guo, Anqi and Zhang, Zhiyuan and Li, Eileen and Gu, Kunlong and Rosenberg, Charles and Zhai, Andrew},
  booktitle={Proceedings of the 26th ACM SIGKDD International Conference on Knowledge Discovery \& Data Mining},
  pages={3203--3212},
  year={2020}
}

@inproceedings{regan2023semi,
  title={Semi-Automated Music Catalog Curation Using Audio and Metadata.},
  author={Regan, Brian and Hristova, Desislava and Beguerisse-D{\'\i}az, Mariano},
  booktitle={ISMIR},
  pages={605--611},
  year={2023}
}

@inproceedings{ji2024lbsc,
  title={Lbsc: A cost-aware caching framework for cloud databases},
  author={Ji, Zhaoxuan and Xie, Zhongle and Wu, Yuncheng and Zhang, Meihui},
  booktitle={2024 IEEE 40th International Conference on Data Engineering (ICDE)},
  pages={4911--4924},
  year={2024},
  organization={IEEE}
}

@inproceedings{mukherjee2023towards,
  title={Towards optimizing storage costs on the cloud},
  author={Mukherjee, Koyel and Shah, Raunak and Saini, Shiv and Singh, Karanpreet and Kesarwani, Harsh and Barnwal, Kavya and Chauhan, Ayush and others},
  booktitle={2023 IEEE 39th International Conference on Data Engineering (ICDE)},
  pages={2919--2932},
  year={2023},
  organization={IEEE}
}

@inproceedings{sanyal2026klas,
  title={KLAS: Using Similarity to Stitch Neural Networks for Improved Accuracy-Efficiency Tradeoffs},
  author={Sanyal, Debopam and Iyer, Anantharaman S and Khare, Alind and Jain, Trisha and Jajoo, Akshay and Lee, Myungjin and Kerce, James Clayton and Tumanov, Alexey},
  booktitle={The Fourteenth International Conference on Learning Representations},
  year={2026}
}

@article{sanyal2023pareto,
  title={Pareto-secure machine learning (PSML): Fingerprinting and securing inference serving systems},
  author={Sanyal, Debopam and Hung, Jui-Tse and Agrawal, Manav and Jasti, Prahlad and Nikkhoo, Shahab and Jha, Somesh and Wang, Tianhao and Mohan, Sibin and Tumanov, Alexey},
  journal={arXiv preprint arXiv:2307.01292},
  year={2023}
}
